\documentclass[
reprint,
amsmath,
amssymb,
aps,
pre,
]{revtex4-2} 

\usepackage{graphicx}
\usepackage{dcolumn}
\usepackage{bm}
\usepackage[hidelinks]{hyperref}

\usepackage{xcolor}
\usepackage{overpic}
\usepackage{comment}
\usepackage{dsfont}
\usepackage{physics}
\usepackage{listings}
\usepackage{enumitem}

\usepackage{cleveref}
\crefname{equation}{Eq.}{Eqs.}
\Crefname{equation}{Equation}{Equations}
\crefname{figure}{Fig.}{Figs.}
\Crefname{figure}{Figure}{Figures}
\crefname{table}{Table}{Tables}
\Crefname{table}{Table}{Tables}
\crefname{section}{Sec.}{Secs.}
\Crefname{section}{Section}{Sections}
\crefname{appendix}{Appendix}{Appendices}
\Crefname{appendix}{Appendix}{Appendices}

\newcommand{\ee}{\mathrm{e}}
\newcommand{\ii}{\mathrm{i}}

\newcommand{\crefs}[1]{Figs.~\ref{#1}}

\newcommand{\asection}[1]{\section{\uppercase{#1}}}

\begin{document}
	\title{Crystal formation in systems of pseudo-forced swarmalators} 
	
	\author{Brennan J. H. Hughes}
	\author{Christoph Bruder}
	\author{Tobias Kehrer}
	\affiliation{Department of Physics, University of Basel, Klingelbergstrasse 82, CH-4056 Basel, Switzerland}
	
	\date{\today}
	
	\begin{abstract}
		Swarmalators are active agents that move in position space and exhibit internal degrees of freedom.
		Due to interactions of their positions and phases of oscillation, they show on the one hand swarming, similar to the effect of flocking of birds. 
		In addition, they exhibit synchronization behavior, analogous to what has been observed in fireflies.
		Previous works studied scenarios in which the phases are forced externally. 
		Here, we consider a pseudo-force that acts on the positions of the swarmalators.
		Due to the resulting attraction towards the center of position space, transitions from the splintered and active phase-wave state to the static antisynchronized state are found.
		To quantify the crystal order of swarmalators, we introduce an order parameter that is based on the Fourier transform of their positions.
	\end{abstract}
	
	\maketitle

	\noindent

	\section{Introduction}
	Swarmalators \cite{OKeeffe2017,sar2025} are a kind of active matter that has attracted a great deal of attention recently.
	They exhibit position and phase-like degrees of freedom.
	Swarmalators show both classical synchronization \cite{Kuramoto1975,Synch_Strogatz,Synch_Pikovsky}, the alignment of features like frequency and phase of oscillators, and swarming \cite{PhysRevLett.75.1226,Carrillo2010,Toner2024}, the spatial ordering of active agents.
	In the last years, swarmalators have been studied in several variations including the number of spatial and phase dimensions \cite{PhysRevE.98.022203,PhysRevE.105.014211,PhysRevE.109.044212}.
	These models are related to e.g., self-assembly of magnetic colloids \cite{Yan2012}, collective frog choruses \cite{Aihara2014}, or vortex arrays of sperm cells \cite{Riedel2005}.
	Realizations include robots \cite{Barcis2019,novkoski2025} and drones \cite{quinn2025}.
	
	In the 2D case, many kinds of interactions have been considered, e.g., finite-cutoff \cite{Lee_2021}, disordered \cite{PhysRevE.104.044214}, and delayed interactions \cite{PhysRevE.109.014205}.
	One type of modified interactions can be interpreted as external seeds that are forcing the phases of swarmalators \cite{Lizarraga2020,PhysRevE.110.054205,PhysRevE.111.044207} or act as a predator-like contrarian \cite{PhysRevE.111.014209}.
	In this work, we focus on forcing that acts on the position of swarmalators only.
	Since the equations of motion are of first order, the additional term is not a physical force; therefore, we use the term \textit{pseudo-force}.
	The stronger the pseudo-force that pulls each swarmalator towards the center, the smaller the swarms (disks or annuli).
	As a result, splintered and active phase-wave states switch to static antisynchronized states.
	Before the point of transition is reached, the number of groups in steady-state examples of the splintered phase wave decreases.
	
	To indicate the transition, we make use of the standard Kuramoto and phase-position-correlation order parameters.
	In addition, inspired by studies of crystal structure in swarmalators \cite{PhysRevE.109.014205,kumpeerakij2025}, we define an order parameter based on the spatial Fourier transform of the swarmalator positions.
	The crystal order of all steady-state classes is increased by the pseudo-force and changes significantly at the transitions of the splintered and active phase wave.
	
	The paper is structured as follows.
	In \cref{sec:model}, we define the pseudo-force extension of the swarmalator model and the relevant order parameters. 
	In the beginning of \cref{sec:transitions}, we present an analytic solution for the steady state of three swarmalators.
	Afterwards, we discuss the behavior of the static synchronized and unsynchronized states as well as the static, splintered, and active phase-wave states.
	We conclude in \cref{sec:conclusion}.

	\section{Model}\label{sec:model}
	The standard model of swarmalators \cite{OKeeffe2017} consists of an equation of motion for the 2D position $\vec{x}_j$ of the $j$th swarmalator and another equation of motion for its phase of oscillation $\theta_j$,
	\begin{align}
	\dot{\vec{x}}_j =&~ \vec{v}_j + \frac{1}{N}\sum_{i\neq j}\left( \frac{\vec{x}_i - \vec{x}_j}{|\vec{x}_i - \vec{x}_j|} (A + J\cos(\theta_i - \theta_j)) \right.\nonumber\\
	& \left. - B \frac{\vec{x}_i - \vec{x}_j}{|\vec{x}_i - \vec{x}_j|^2}\right)\,,\label{eq:standard_xdot}\\
	\dot{\theta}_j =&~ \omega_j + \frac{K}{N}\sum_{i\neq j} \frac{\sin(\theta_i - \theta_j)}{|\vec{x}_i - \vec{x}_j|}\,.\label{eq:standard_thatadot}
	\end{align}
	In the following, we choose to study identical swarmalators in the frame of the center of mass movement $\vec{v}_j=\vec{v}\to (0,0)$ and in the rotating frame of the mean oscillation frequency $\omega_j=\omega\to 0$.
	Furthermore, we use the standard rescaling of space and time coordinates to unitless quantities and set $A=B=1$.
	The remaining parameter space $(K,J)$ consists of five classes of states, see \cref{tab:standard_phases} below and Fig.~1 of \cite{OKeeffe2017}.
	These phases are called static synch (StS), static asynch (StAS), static phase wave (StPW), splintered phase wave (SpPW), and active phase wave (ActPW).
	The first three steady states are static in time, whereas the latter two steady states feature active movement in position space and varying phases of oscillation of swarmalators.
	
	\begin{table}[b]
		\centering
		\begin{tabular}{lcc}
			\hline\hline
			\textbf{State} & $K$ & $J$\\\hline
			Static Synch (StS) & $1$ & $0.1$\\
			Static Asynch (StAS) & $-1$ & $0.1$\\
			Static Phase Wave (StPW) & $0$ & $1$\\
			Splintered Phase Wave (SpPW) & $-0.1$ & $1$\\
			Active Phase Wave (ActPW) & $-0.75$ & $1$\\\hline\hline
		\end{tabular}
		\caption{List of standard swarmalator phases shown in the top left panels of \crefs{fig:crystal_StS_StAS}(a), \ref{fig:crystal_StS_StAS}(b), \ref{fig:crystal_StPW}, \ref{fig:crystal_SpPW_Act_PW}(a), and \ref{fig:crystal_SpPW_Act_PW}(b).
		}
		\label{tab:standard_phases}
	\end{table}

	\subsection{Pseudo-forced swarmalators}
	In previous works, forced swarmalators have been studied in a scenario where an additional term proportional to $F \cos(\omega_s t - \theta_j)$ has been added to the equation of motion of the phase $\theta_j$ \cite{Lizarraga2020,PhysRevE.110.054205,PhysRevE.111.044207}.
	This term can be interpreted as the phase locking of all swarmalators to a seed-like object that oscillates with a frequency $\omega_s$.
	
	In contrast, in this work, we consider a pseudo-force $\lambda (\vec{x}_0 - \vec{x}_j)$ acting on the positions of the swarmalators. 
	The parameter $\lambda$ controls the strength of the pseudo-force.
	Using the assumptions mentioned below \cref{eq:standard_thatadot}, we define
	\begin{align}
	\dot{\vec{x}}_j =&~ \frac{1}{N}\sum_{i\neq j}\left( \frac{\vec{x}_i - \vec{x}_j}{|\vec{x}_i - \vec{x}_j|} (1 + J\cos(\theta_i - \theta_j)) - \frac{\vec{x}_i - \vec{x}_j}{|\vec{x}_i - \vec{x}_j|^2}\right)\nonumber\\
	& + \lambda (\vec{x}_0 - \vec{x}_j)\,,\label{eq:extension_xdot}\\
	\dot{\theta}_j =&~ \frac{K}{N}\sum_{i\neq j} \frac{\sin(\theta_i - \theta_j)}{|\vec{x}_i - \vec{x}_j|}\,, \label{eq:extension_phidot}
	\end{align}
	where we set $\vec{x}_0=(0,0)$ equal to the origin of the 2D position space.
	The equation of motion of the center of mass $\vec{x}_c = \sum_j \vec{x}_j/N$,
	\begin{align}
	\dot{\vec{x}}_c = \frac{1}{N}\sum_j \dot{\vec{x}}_j = \lambda (\vec{x}_0 - \vec{x}_c)\,,
	\end{align}
	is useful to interpret $\lambda$: it is the decay rate of the center of mass towards the center of attraction of the pseudo-force.
	In all simulations discussed in the following, we draw the initial positions of $N=300$ swarmalators from a uniform distribution $\vec{x}_j\in[0,2]\times[0,2]$ and subtract their mean afterwards.
	This results in a distribution of swarmalators with constant center of mass $\vec{x}_c =(0,0)$.
	The distribution of the initial phases is uniform on $[-\pi,\pi]$.

	\subsection{Order parameters}
	To classify steady states, we use the Kuramoto order parameter 
	\begin{align}
	R = \frac{1}{N}\sum_{j=1}^N \ee^{\ii \theta_j}\,,
	\end{align}
	and the order parameter of phase-position correlation $S=|W|$ \cite{OKeeffe2017}, where
	\begin{align}
	W &= S\,\ee^{\ii \Psi} = \begin{cases}W_+ & \text{for~}|W_+| > |W_-| \\ W_- & \text{for~}|W_+| < |W_-|\end{cases}\,,\label{eq:S_W_ord_par}\\
	W_\pm &= \frac{1}{N}\sum_{j=1}^N \ee^{\ii (\phi_j \pm \theta_j)}\,.
	\end{align}
	Here, $\theta_j$ is the phase of the $j$th swarmalator and $\phi_j$ is the azimuthal angle of its 2D position.
	The order parameter $|R|$ indicates the global phase synchronization of all swarmalators, whereas $S$ indicates whether the value of the phase of a swarmalator (anti)correlates with its 2D position. 
	
	To characterize crystal order, i.e., the main contribution of this work, we define an additional order parameter
	\begin{align}
	\mathcal{F}_\mathrm{max} = \max_{\vec{k}\in \mathcal{K}}|\mathcal{F}(\vec{k})|\,,\label{eq:Fmax}
	\end{align}
	which is based on the Fourier transform of the density distribution $\rho(\vec{x})=\sum_{j=1}^N \delta(\vec{x}-\vec{x}_j)/N$,
	\begin{align}
	\mathcal{F}(\vec{k}) = \frac{1}{N}\sum_{j=1}^N \exp(-\ii\vec{k}\vec{x}_j)\,.
	\label{eq:FFT_2D}
	\end{align}
	In the numerical calculation of $\mathcal{F}_\mathrm{max}$, we consider $\vec{k}\in\mathcal{K}$ where we exclude a region close to the center of momentum space at which $\mathcal{F}(\vec{0})=1$.

	\section{Induced Phase Transitions}\label{sec:transitions}
	In this section, we will see that with increasing pseudo-force parameter $\lambda$ states from all classes become more dense and show higher crystal order.
	In particular, states from the classes SpPW and ActPW undergo a transition to StAS.
	Another transition, the formation of the annulus in StPW, is discussed in Appendix~\ref{sec:appendix_StPW}.

	\begin{figure}[t]
		\centering
		\includegraphics[width=8.5cm]{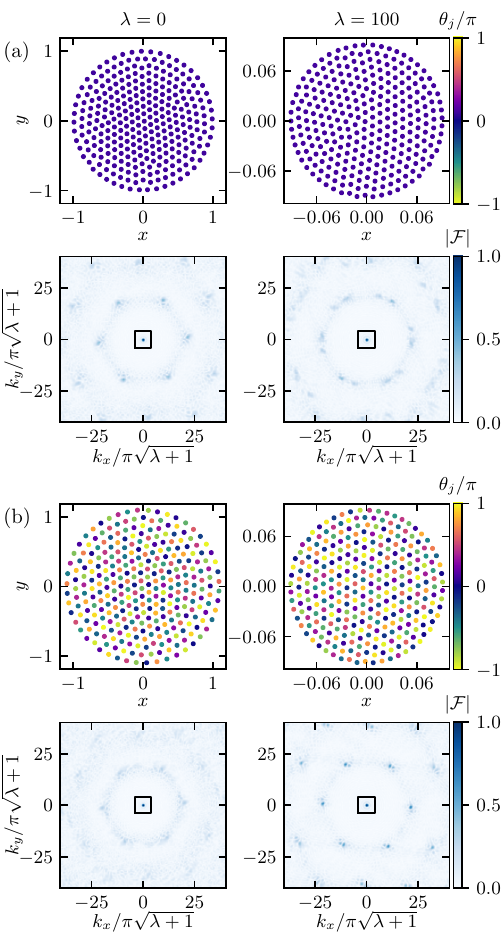}
		\caption{Examples of StS and StAS steady states defined in Table~\ref{tab:standard_phases} at time $t=10^4$ for $\lambda=0,100$ and corresponding Fourier transforms $\mathcal{F}$ of the positions, see \cref{eq:FFT_2D}.
			The larger $\lambda$, the denser the group of swarmalators (note the change of scale).
			(a) StS for $(K,J)=(1,0.1)$.
			(b) StAS for $(K,J)=(-1,0.1)$.
			The black squares in the plots of the Fourier transforms correspond to the excluded region when computing $\mathcal{F}_\text{max}$, see \cref{eq:Fmax}.
		}
		\label{fig:crystal_StS_StAS}
	\end{figure}

	\subsection{Analytic stable configurations}\label{sec:analytics}
	In the simple example of three swarmalators, we can find a fixed-point solution to the equations of motion.
	This three-swarmalator scenario can be described by three distances $d_{ij}=|\vec{x}_i - \vec{x}_j|$ and three phase differences $\theta_{ij}=\theta_i-\theta_j$.
	The resulting equations of motion for the phase difference $\theta_{ij}$ read
	\begin{align}
	\dot{\theta}_{ij} &= \frac{K}{N}\left(\sum_{n\neq i}\frac{\sin(\theta_{ni})}{d_{ni}}  - \sum_{m\neq j}\frac{\sin(\theta_{mj})}{d_{mj}} \right)\nonumber\\
	&= -\frac{2K}{N d_{ij}}\sin(\theta_{ij}) +\frac{K}{N} \sum_{k\neq i,j}\left(\frac{\sin(\theta_{ki})}{d_{ki}} - \frac{\sin(\theta_{kj})}{d_{kj}} \right)\,.
	\end{align}
	In the case of $N=3$ swarmalators, there are two free parameters $\theta_{12}$ and $\theta_{23}$ since $\theta_{13}=\theta_{12}+\theta_{23}$.
	One steady state for $\dot{\theta}_{12}=\dot{\theta}_{23}=0$ can be found by setting $d_{ij}=d$ leading to the stable solution $\theta_{ij}=0$ ($\theta_{ij}=\pm 2\pi/3$) for $K>0$ ($K<0$).
	This choice can be interpreted as an approximation of one equilateral triangular unit cell of the crystals we will discuss in the remainder of this section.
	The differential equation of the time evolution of the distances
	\begin{align}
	\dot{d}_{ij} = \frac{(\dot{\vec{x}}_i - \dot{\vec{x}}_j)(\vec{x}_i - \vec{x}_j)}{|\vec{x}_i - \vec{x}_j|}\,,
	\end{align}
	for $K>0$ and $\theta_{ij}=0$ reduces to 
	\begin{align}
	\dot{d} = \frac{1}{2} - \frac{1+J}{2} d - \frac{\lambda}{3}d^2\,.
	\end{align}
	The steady state distance is given by
	\begin{align}
	d_\text{S} = \frac{\sqrt{9(1+J)^2 + 24\lambda}-3(1+J)}{4\lambda} \stackrel{\lambda\to\infty}{\approx}\sqrt{\frac{3}{2\lambda}}\,.
	\end{align}
	In the case of $K<0$ and $\theta_{ij}=\pm 2\pi/3$, we obtain
	\begin{align}
	\dot{d} = \frac{1}{2} - \frac{2-J}{4} d - \frac{\lambda}{3}d^2\,,
	\end{align}
	with steady state distance
	\begin{align}
	d_\text{AS} = \frac{\sqrt{9(2-J)^2 + 96\lambda}-3(2-J)}{8\lambda} \stackrel{\lambda\to\infty}{\approx}\sqrt{\frac{3}{2\lambda}}\,.
	\end{align}
	Note that the distance in the synchronized case is smaller than the distance in the antisynchronized case: $d_\text{S} \leq d_\text{AS}$ with equality if $J=0$.
	
	These results are a first hint for the $1/\sqrt{\lambda}$ scaling of the size of the steady states for large $\lambda$.
	In the case of a linear attraction $\lambda (\vec{x}_i - \vec{x}_j)$ between swarmalators, the resulting radius in the synchronized StS case is proportional to $1/\sqrt{\lambda + 1}$ \cite{OKeeffe2017}.
	As we will see in \cref{sec:StS_StAS}, this approximation suits all classes of steady states.

	\subsection{Static synch and static asynch}\label{sec:StS_StAS}
	The two classes StS and StAS lie at the two ends of the scenario $|K|\gg J$.
	In this limit, if $K>0$, all swarmalators synchronize to the same phase $\theta_j=\theta$ leading to $|R|=1$.
	In contrast, if $K<0$, the swarmalators tend to antisynchronize with each other such that their phases become uniformly distributed and $|R|=0$.
	
	In \cref{fig:crystal_StS_StAS}, we show steady-state examples for $\lambda=0,100$ and their corresponding Fourier transforms $\mathcal{F}(\vec{k})$.
	The rescaling of $\vec{k}$ with $1/\sqrt{\lambda+1}$, see the discussion at the end of \cref{sec:analytics}, approximately counteracts the rescaling of the distances due to increasing $\lambda$.
	The steady-state pictures show triangular lattices and the corresponding $\mathcal{F}$ exhibit isolated local maxima.
	The higher the crystal order, the more pronounced the local maxima.
	In the cases where the first ``ring of local maxima" exhibits more than six spots, multiple domains can be identified in the corresponding steady-state picture.
	For example, in \cref{fig:crystal_StS_StAS}(a) at $\lambda=100$ \textit{two} domains are visible (a smaller one in the top left corner) leading to six \textit{pairs} of local maxima in the plot of the corresponding Fourier transform. 
	The black squares in the plots of the Fourier transforms correspond to the excluded region when computing $\mathcal{F}_\text{max}$, see \cref{eq:Fmax}.

	\begin{figure}[t]
		\centering
		\includegraphics[width=8.5cm]{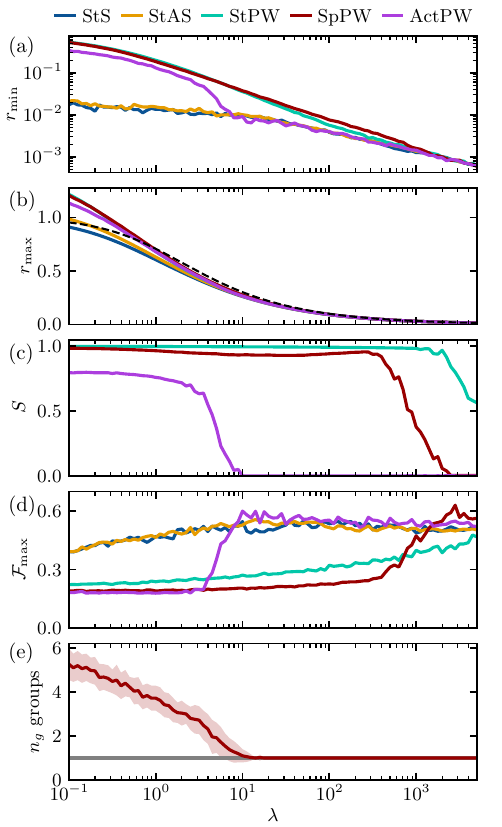}
		\caption{Various quantities characterizing steady states of $N=300$ swarmalators averaged over $50$ realizations.
			(a) The smallest distances $r_\text{min}$ of any swarmalator from the origin.
			(b) The largest distances $r_\text{max}$ of any swarmalator from the origin.
			The black dashed curve corresponds to $1/\sqrt{\lambda+1}$, see end of \cref{sec:analytics}.
			(c) Order parameter $S$ of phase-position correlation, see \cref{eq:S_W_ord_par}.
			(d) Fourier-transform order parameter $\mathcal{F}_\text{max}$ quantifying crystal order, see \cref{eq:Fmax}.
			(e) Number of groups $n_g$ in SpPW class, see Appendix~\ref{sec:number_of_groups} for a detailed description of the method.
			The shaded region corresponds to one standard deviation around the mean and the gray horizontal line indicates $n_g=1$.
			The $(K,J)$ parameters for each class of steady states are listed in \cref{tab:standard_phases}.
		}
		\label{fig:crystal_order_parameters}
	\end{figure}
	
	In \cref{fig:crystal_order_parameters}, we present the data for $N=300$ swarmalators averaged over $50$ realizations for all classes of steady states. 
	The scaling of the smallest distance $r_\text{min}$ of any swarmalator from the origin is shown in \cref{fig:crystal_order_parameters}(a).
	The $r_\text{min}$ of StS and StAS can be used as baselines to distinguish disc-like position distributions from annulus-like ones.
	The scaling of the largest distance $r_\text{max}$ of any swarmalator from the origin is shown in \cref{fig:crystal_order_parameters}(b). 
	The black dashed curve corresponds to $1/\sqrt{\lambda+1}$. 
	In \cref{fig:crystal_order_parameters}(d), the order parameter $\mathcal{F}_\text{max}$ shows a small increase for increasing $\lambda$ corresponding to higher crystal order.

	\begin{figure*}[t]
		\centering
		\includegraphics[width=17cm]{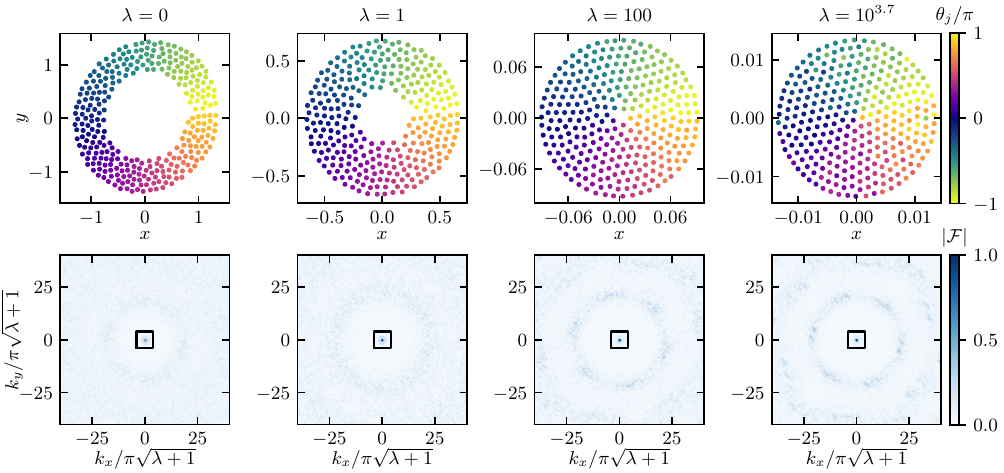}
		\caption{Examples of StPW steady states for $(K,J)=(0,1)$ at time $t=2000$ for different values of $\lambda$ and corresponding Fourier transforms of the position, see \cref{eq:FFT_2D}.
			The larger $\lambda$, the denser the annulus of swarmalators and the more crystal-like its structure.
			The black squares in the plots of the Fourier transforms correspond to the excluded region when computing $\mathcal{F}_\text{max}$, see \cref{eq:Fmax}.
		}
		\label{fig:crystal_StPW}
	\end{figure*}

	\subsection{Static phase wave}\label{sec:StPW}
	In the remaining class of static steady states, StPW at $(K,J)=(0,1)$, we can see in \cref{fig:crystal_order_parameters}(a) that $r_\text{min}$ is larger than the one of StS and StAS due to the annulus-like structure of StPW.
	Another consequence of this structure is that the crystal order is small and increases slowly with increasing $\lambda$, see \cref{fig:crystal_order_parameters}(d).
	For large $\lambda > 1000$, $S$ decreases indicating a possible transition to StAS.
	In \cref{fig:crystal_StPW}, we present steady-state examples for various values of $\lambda$ and their corresponding Fourier transforms $\mathcal{F}(\vec{k})$.
	The larger $\lambda$, the smaller the inner radius of the annulus and the higher the crystal order.
	In Appendix~\ref{sec:appendix_StPW}, we apply the Fourier-transform order parameter to another scenario: at $\lambda=0$ and varying the standard deviation of the initial distribution of phases leads to a transition between a disk-like and annulus-like spatial distribution.

	\subsection{Active phases}\label{sec:SpPW_ActPW}
\begin{figure*}[t]
		\centering
		\includegraphics[width=17cm]{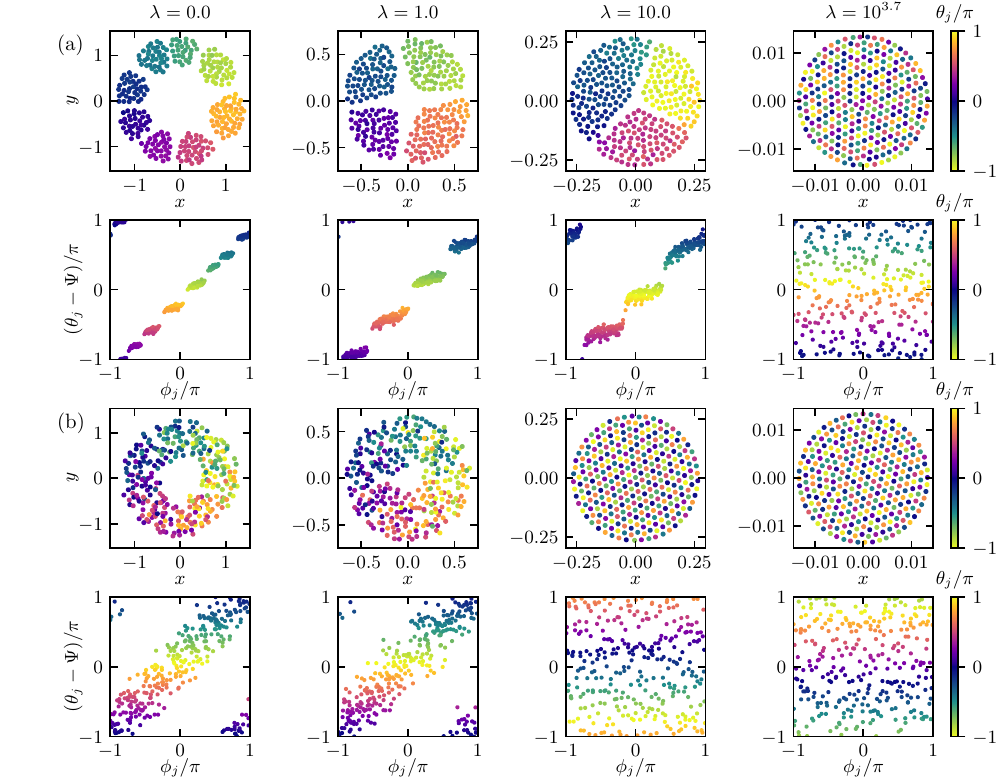}
		\caption{Examples of SpPW and ActPW steady states at time $t=500$ for different values of $\lambda$ and corresponding phase-position correlation, see \cref{eq:S_W_ord_par}.
			The larger $\lambda$, the smaller the number of groups in the SpPW phase and the more crystal-like the structure in the ActPW phase.
			(a) SpPW for $(K,J)=(-0.1,1)$.
			(b) ActPW for $(K,J)=(-0.75,1)$.
		}
		\label{fig:crystal_SpPW_Act_PW}
	\end{figure*}

	Both SpPW and ActPW are classes of active steady states.
	Swarmalators in the SpPW class perform movements in separated groups that are aligned on an annulus around the origin of position space.
	The center of mass of these groups is almost static.
	In contrast, swarmalators in the ActPW class perform larger movements around the origin of position space within an annulus.
	In both cases, increasing $\lambda$ leads to a closing of hole of the annulus and a transition to StAS.
	Examples of steady states and corresponding plots of the correlation between phase and position are presented in \cref{fig:crystal_SpPW_Act_PW}.
	The phases $\theta_j$ are shifted by the complex argument $\Psi$ of the order parameter $W$, see \cref{eq:S_W_ord_par}, and the position information is reduced to the azimuthal angle $\phi_j$.
	
	\begin{figure}[!t]
		\centering
		\includegraphics[width=8.5cm]{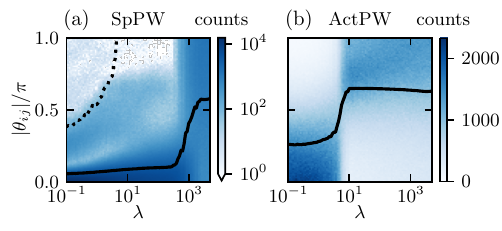}
		\caption{Histogram of phase difference $|\theta_{ij}|$ ($101$ bins) between swarmalators on the same triangular unit cell (triangulation) for each value of $\lambda$ taking into account $50$ realizations.
			(a) SpPW for $(K,J)=(-0.1,1)$. 
			The dotted curve corresponds to $|\theta_{ij}| = 2\pi/n_g$, where $n_g$ is the number of groups, see \cref{fig:crystal_order_parameters}(e).
			(b) ActPW for $(K,J)=(-0.75,1)$.
			In both panels, the solid curve corresponds to the mean absolute phase difference.
		}
		\label{fig:phase_diff_SpPW_Act_PW}
	\end{figure}

	In \cref{fig:crystal_order_parameters}(a), we see that the inner radius $r_\text{min}$ of the annuli of StPW and SpPW behave similar, whereas $r_\text{min}$ of ActPW drops rapidly when the transition to StAS occurs.
	The transition from SpPW and ActPW to StAS is visible in both \crefs{fig:crystal_order_parameters}(c) and \ref{fig:crystal_order_parameters}(d), where $S$ and $\mathcal{F}_\text{max}$ are presented.
	The phase-position correlation drops to $S=0$ and the crystal order $\mathcal{F}_\text{max}$ increases.
	
	Moreover, the number $n_g$ of groups in the SpPW class decreases for increasing $\lambda$, see \crefs{fig:crystal_order_parameters}(e) and \ref{fig:crystal_SpPW_Act_PW}(a).
	We use a method described in more detail in Appendix~\ref{sec:number_of_groups} to quantify the number of groups.

	In \cref{fig:phase_diff_SpPW_Act_PW}, we present another way to visualize the transition from SpPW and ActPW to StAS.
	We calculate the phase difference $|\theta_{ij}|$ between neighboring swarmalators to distinguish between phase synchronization $|\theta_{ij}|=0$ and antisynchronization $|\theta_{ij}|=2\pi/3$. 
	Here, neighbors are defined as the swarmalators that are located at the corners of one triangular unit cell obtained by Delaunay triangulation \cite{VandenHeuvel2024DelaunayTriangulation}.
	For each triangle, we therefore get three phase differences.
	Taking all $50$ realizations into account, we create a histogram for several values of $\lambda$ with $101$ bins in $|\theta_{ij}|$.
	The black solid curves in \cref{fig:phase_diff_SpPW_Act_PW} correspond to the mean phase differences indicating a transition similar to \crefs{fig:crystal_order_parameters}(c) and \ref{fig:crystal_order_parameters}(d).
	The dotted curve in \cref{fig:phase_diff_SpPW_Act_PW}(a) corresponds to $|\theta_{ij}|=2\pi/n_g$, where $n_g$ is the number of groups in a SpPW steady state, see \cref{fig:crystal_order_parameters}(e).
	This curve is close to the second local maximum of the histogram that can be interpreted as the phase difference between swarmalators that lie on a triangle that connects two different groups.

	\section{Conclusion}\label{sec:conclusion}
	In this work, we have considered swarmalators that are subject to an additional linear pseudo-force acting on their 2D position.
	This pseudo-force pulls the swarmalators towards the center of position space and induces transitions from the active phase-wave state and splintered phase-wave state to the static antisynchronized state.
	Since the swarms of every steady-state class become more dense with increasing pseudo-force, additional indicators of the transition have been found: the number of groups in the splintered phase wave class decreases and the crystal order increases.
	The latter effect we have quantified by proposing an order parameter that is based on the spatial Fourier transform of the swarmalators.
	If the swarm aligns in a highly-ordered triangular crystal, its Fourier transform exhibits a small number of sharp peaks.
	The more domains or the less order, the height of these peaks decreases.
	Using Delaunay triangulation, we have studied the phase difference between neighboring swarmalators as another order parameter of the transitions from active phase-wave states and splintered phase-wave states to static antisynchronized states.
	
	The Fourier-transform order parameter can be beneficial for studies of crystal order in other swarmalator setups.
	In the future, other implementations of forcing can be realized: (i) pseudo-force with different distance scaling, (ii) multiple pseudo-forces pulling towards different locations in position space, or (iii) combination of forces acting on position and phase inducing frustration.
	Moreover, the phase diagram $(K,J,\lambda)$ can be explored for $J>1$ since $\lambda$ might counteract potential repulsion between swarmalators.

	\section*{Acknowledgments}
	We thank Julian Arnold, Tobias Nadolny, Frank Schäfer, and Benjamin Senn for fruitful discussions.
	T.K. and C.B.~acknowledge financial support from the Swiss National Science Foundation individual grant (Grant No.~200020 200481). 
	We furthermore acknowledge the use of \textit{\hbox{DifferentialEquations.jl}} \cite{rackauckas2017differentialequations}, \textit{\hbox{DelaunayTriangulation.jl}} \cite{VandenHeuvel2024DelaunayTriangulation}, and \textsc{OpenCV} \cite{opencv_library}.
	The data that support the findings of this article are openly available \cite{data}.

	\appendix
	
	\makeatletter
	\renewcommand\appendixname{APPENDIX}
	\renewcommand{\@hangfrom@section}[2]{
		\@hangfrom{#1}\MakeUppercase{#2: }
	}
	\makeatother

	\asection{Number of groups}\label{sec:number_of_groups}
	In this appendix, we describe the algorithm we used for finding the number of groups in steady states of the SpPW class.
	The \textsc{Python} code is provided in \cite{data}.
	Here, we list key elements: 
	\begin{enumerate}[leftmargin=5pt, itemindent=\parindent, labelwidth=0pt]
		\item First, we convert the position data into a \textit{matplotlib} object and then into an array \textit{img} that can be processed by \textsc{OpenCV}.
		We then use \textit{cv.cvtColor(img, cv.COLOR\_BGR2GRAY)} to turn the image into a grayscale image and set pixel values above $200$ to $255$.
		\item In the next step, we perform a so-called distance transform. 
		The value of each pixel is replaced by $255$ if the distance to the nearest zero pixel is larger than the threshold distance $p_0$.
		This is one parameter of the algorithm.
		Now, the dot for each swarmalator has effectively increased its size such that (ideally only) swarmalators of the same group overlap.
		\item Next, we use \textit{cv.findContours()} to find the contour of each group.
		We discard groups that have an area that is smaller than a minimum value parametrized by the second parameter $p_1$.
		\item Finally, we calculate the convex hull of these contours.
		The number of groups $n_g=\max\lbrace 1,n_h\rbrace$ calculated by our algorithm is defined as the number of hulls $n_h$ or 1 if $n_h=0$.
	\end{enumerate}
	
	\begin{figure}[t]
		\centering
		\includegraphics[width=8.5cm]{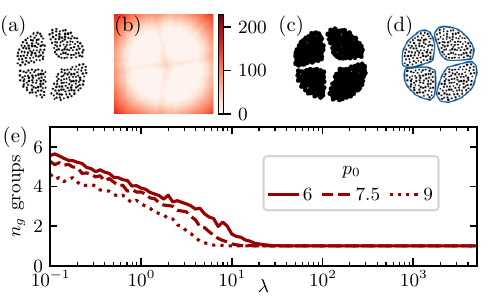}
		\caption{Intermediate steps of the group-finding algorithm for $N=300$ swarmalators in the SpPW phase at $(K,J)=(-0.1,1)$.
			(a) Initial image of swarmalator positions.
			Black (white) pixels corresponds to a value of $0$ ($255$).
			(b) Distance transform: the value of each pixel is replaced by its distance to the nearest zero pixel.
			(c) The pixel value is set to $255$ if the distance to the next zero pixel is above a threshold distance $p_0$. 
			Here, we set $p_0=7.5$. 
			(d) Finding contours and discarding the ones that enclose too few swarmalators.
			The blue surrounding curves correspond to the convex hulls of the contours.
			(e) Number of groups $n_g$ for different values of $p_0$.
		}
		\label{fig:group_alg_example}
	\end{figure}

	In \cref{fig:group_alg_example}, we show an example for the different steps of the algorithm.
	The resulting $n_g$ for different values of $p_0$ is shown in \cref{fig:group_alg_example}(e).
	However, the qualitative result that $n_g$ decreases approximately logarithmically with $\lambda$ is unchanged.

	\begin{figure}[t]
		\centering
		\includegraphics[width=8.5cm]{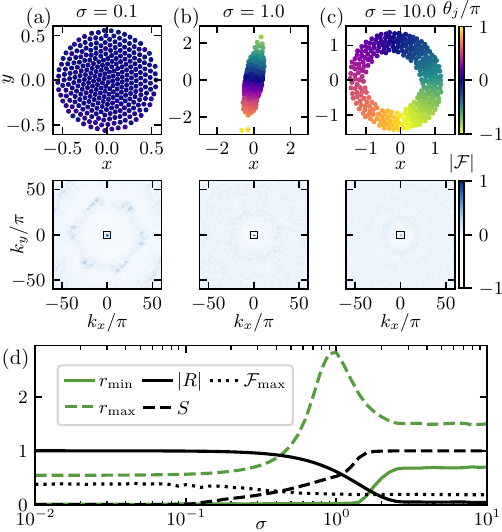}
		\caption{Annulus formation for $N=300$ swarmalators in the StPW phase at $(K,J)=(0,1)$ and $\lambda=0$.
			(a)--(c) Examples of steady states for selected values of the standard deviation $\sigma$ of the phase distribution with corresponding Fourier transforms of the positions, see \cref{eq:FFT_2D}.
			(d) Order parameter values averaged over $50$ realizations.
		}
		\label{fig:annulus_formation}
	\end{figure}

	\asection{Annulus formation}\label{sec:appendix_StPW}
	Another example to which the Fourier-transform order parameter can be applied is the formation of the annulus in the StPW class for different standard deviations of the initial phases at $\lambda=0$.
	Since $K=0$ in StPW, the initial distribution of phases does not evolve in time.
	
	Here, we draw the initial phases from a normalized Gaussian distribution with zero mean and standard deviation $\sigma$. 
	In \crefs{fig:annulus_formation}(a) to \ref{fig:annulus_formation}(c), we present steady-state examples for selected values of $\sigma$ and their corresponding spatial Fourier transforms, see \cref{eq:FFT_2D}.
	The order parameters $|R|$, $S$, $r_\text{min}$, $r_\text{max}$, and $\mathcal{F}_\text{max}$ indicate the transition from disk-like to annulus-like spatial distributions, see \cref{fig:annulus_formation}(d).
	If $\sigma \ll 1$, the steady state is disk-shaped with significant crystal order, the phases are almost identical ($|R|\approx 1$) and there is no correlation between phase and position ($S\approx0$).
	If $\sigma \gg 1$, the swarmalators are distributed on an annulus with low crystal order, see \cref{sec:StPW}.
	The phases are almost evenly distributed ($|R|\approx 0$) and there is a strong phase-position correlation ($S \approx 1$).

	\clearpage

\end{document}